# Evaluation and comparison of NiO/CuO nanocomposite and NiO nanofibers as non-enzymatic electrochemical cholesterol biosensor

Zahra Gholami Shiri, Seyed Mojtaba Zebarjad, Kamal Janghorban

**Abstract**

Fabrication of a highly sensitive non-enzymatic electrochemical sensor for the detection of cholesterol is a challenging issue due to its low redox activity. In the present study, the electrochemical response of a novel combination of metal oxides has been explored. NiO/CuO nanocomposite was fabricated by electrospinning method and its morphology and structural characteristics were studied using scanning electron microscopy, EDX and X-ray diffraction techniques. Electrochemical studies revealed that the NiO/CuO/GC electrode poses a linear detection range between 0.8 to 6.5 mM, low detection limit of 5.9 µM and a sensitivity of 10.27 µA.mM$^{-1}$.cm$^{-2}$ in the neutral buffer solution. The fabricated electrode has a response time of 3 seconds and a good ability to discriminate interfering species. Moreover, the nanocomposite electrode exhibited high stability as well as the other favorable factors, makes it an effective sensor for the detection of cholesterol. In addition, cyclic voltammetry experiments showed that the current response of NiO/GCE is approximately 30 % of the NiO/CuO/GCE response, suggesting that the nanocomposite has a greater electrocatalytic activity towards cholesterol oxidation than NiO nanofibers. NiO/GCE results showed the sensitivity of 2.5 µA.mM$^{-1}$.cm$^{-2}$, response time of 2 seconds and a limited stability over three-week time span.

**Introduction**

Cholesterol is a steroid compound and a constituent of human cell membrane. The normal level of cholesterol is 3.1- 6.7 mM. An increase in concentration of cholesterol leads to cardiovascular diseases, as its deficiency causes other disorders like anemia, hyperthyroidism, malabsorption, and so on [1]. Accordingly, regular monitoring of cholesterol level of body serum is vital and it stimulates the development of an accurate and economic technique. Among various detection methods such as spectrophotometric, colorimetric and high performance liquid chromatography (HPLC), the highest tendency is toward the electrochemical sensing approach because of its high sensitivity and efficiency [2].

Over the past few decades, the enzymatic approach of cholesterol detection, using cholesterol oxidase (ChOx) and cholesterol esterase (ChEt) has been widely investigated [1-5]. In this method the immobilized enzyme (ChOx) catalyzes the cholesterol oxidation in the presence of oxygen and produces hydrogen peroxide. The well-known equation for this reaction is expressed in the following [5]:

$$\text{Cholesterol} + O_2 \xrightarrow{\text{ChOx}} \text{Cholest}-4-\text{en}-3-\text{one} + H_2O_2. \qquad \text{(Eq. 1)}$$

The enzyme-based sensors demonstrate great selectivity and sensitivity. However, problems associated with the inherent characteristics of enzymes have limited the practical application of this approach. For instance, immobilization process can denature the enzyme. Besides, the activity of enzymes decreases over time and can be affected by temperature, toxic substances, pH and humidity [2]. Enzyme free biosensors that directly oxidize cholesterol in the sample, are an interesting alternative for overcoming the enzymatic biosensors' flaws. Despite practical application of non-enzymatic systems, there are limited efforts for enzyme free detection of cholesterol. This is mostly due to the limited redox activity of cholesterol to specific conditions and far positive/negative potentials [6]. Also, the slow kinetic of analyte oxidation at bare electrode and the limited number of systems operating in physiological pH are other major setbacks in fabrication of non-enzymatic sensors [7]. The geometry of the electrode is a determining factor in the kinetic of the reactions. With the development of nano-dimensioned materials, the problems of overpotentials of electrochemical reactions have been addressed. The electrooxidation process includes the absorbance of the analyte on the surface of the electrode. Nanostructures provide a large surface area and several analyte molecules can be absorbed to the electrode active sites at the same time. As a result, nanomaterials can increases the rate of reaction, improve the electron transfer and decrease the working potential of reactions [7, 8].

In order to fully exploit the surface area, electrospun nanofiber mats with high surface to volume ratio are of great interest. Electrospinning is a robust manufacturing rout that has attracted great attention. The highlight of this method lies in the fabrication simplicity and wide applicability to many materials. Due to its outstanding versatility, electrospinning has found numerous applications ranging from biomedical engineering [9-11] to electronics industry [12] and energy storage [13, 14]. A number of biosensors based on electrospun nanofibers have been reported [15-17]. However, application of electrospun nanofibers to the cholesterol biosensors is far less widespread.

Direct non-enzymatic electrooxidation of cholesterol considerably depends on the electrode material [6]. A number of electrode materials have been used over the past decade. Noble metals nanostructures such as porous silver nanotubes [2], silver nanoparticle [18] and gold electrode modified with platinum nanoparticles [19] have been designed for cholesterol sensing. In addition, organic materials like polyaniline/multiwall carbon nanotube [20], polyaniline/graphen composite [21] and functionalized graphene [22], have been reported for enzyme less detecting of cholesterol.

Transition metal oxides and in particular nickel and copper oxides have received a significant attention in biosensing applications due to their noticeable features such as chemical stability, redox process at different potential ranges and low cost [23-26]. Based on the composition of NiO and CuO with other metal oxide/ organic compounds, several electrochemical sensors have been investigated [27-29]. However, the synergistic effect of these two metal oxides on the electrocatalysis of cholesterol has not been examined.

To the best of our knowledge, there is no available report on the electrochemical enzyme free performance of the nanostructured NiO/CuO composite toward cholesterol detection. Hence, in order to fill the literature gap, NiO/CuO nanocomposite was fabricated by electrospinning technique and characterized by cyclic voltammetry and chronoamperometry methods. Furthermore, electrochemical response of NiO nanofibers toward cholesterol was evaluated and compared to NiO/CuO nanocomposite. The experimental procedure of this research is provided in the following section which will be followed by the results and their discussion. The paper is finished with a brief summary and conclusion of the work.

**Experimental procedure**

*Materials*

The purchased materials for the current study are as follows; Poly vinyl alcohol (PVA, Mw = 72,000, Merck), Nickel acetate tetrahydrate (Ni(OCOCH3)$_2$.4H$_2$O, Sigma-Aldrich), Copper nitrate trihydrate (Cu(NO$_3$)$_2$.3H$_2$O, Sigma), dimethylformamide (DMF, C$_3$H$_7$NO, Merck), Monosodium phosphate (NaH$_2$PO$_4$.2H$_2$O, Merck), Disodium phosphate (Na$_2$HPO$_4$.2H$_2$O,

Merck), Sodium hydroxide (NaOH, Merck) and Nafion ($C_7HF_{13}O_5S.C_2F_4$, Merck). These materials are utilized without any further treatment. In addition, Cholesterol ($OC_{27}H_{46}$), Urea, Ascorbic acid, Tyrosine, Glucose and Glycine were supplied from Sigma Aldrich. The surfactant for the experiments is Triton-X100 ($C_{14}H_{22}O(C_2H_4O)n$, Acros). With the use of double distilled water, the aqueous solutions were prepared.

### *NiO/CuO nanocomposite preparation*

The Electrospinning solution of nickel acetate was prepared by dissolving 0.65 g nickel acetate in 4 ml PVA solution (7%wt) and was mixed for 12 hours. To prepare copper nitrate solution, 0.3 g copper nitrate and 0.425 g PVA were dissolved in 4.5 ml DMF and 4.25 ml DI water, respectively. These two solutions were mixed for 6 hours. The prepared solutions were loaded in 5 ml medical syringes, individually. Electrospinning was carried out by an instrument made by NanoAzma Co, in Iran. The fibers were collected on the rotating drum (constant speed, 100 rpm). The two salt solutions were electrospinned simultaneously with two nozzles with the same process parameters. Electrospinning proceeded under feed rate of 0.2 ml/h, voltage of 14 kV, needle to target distance of 15 cm and at ambient temperature and humidity. The electrospun fibers calcined at 550$^o$C for 5 hours at a heating rate of 5$^o$C/min in an electrical furnace (Mahar Furnace Co, Iran).

### *Electrode preparation*

Glassy carbon electrode surface (3 mm in diameter) was polished, ultra-sonicated in the mixture of aceton and water (1:7) and rinsed with water several times. Four mg of nanocomposite powder was dispersed in 20 μl nafion and 100 μl distilled water for several minutes. The electrode surface was ink-coated by 50 μl of ink using spin coating (vCOAT4-HI, Backer) and it was allowed to dry in the air.

### *Cholesterol solution preparation*

Phosphate buffer solution (PBS 0.1 M) was prepared following the Henderson–Hasselbalch equation (Eq. 2) [30];

pH = pKa + log (conjugate base/acid). (Eq. 2)

To prepare the buffer solution, monosodium phosphate and disodium phosphate were dissolved in distilled water. The pH adjusted to the biological pH of 7.4 by adding NaOH aqueous solution. To prepare cholesterol solution in different concentrations, cholesterol powder was solved in PBS 0.1M under ultrasonication and injecting Triton X100 as surfactant.

*Characterization*

The morphology, elemental map and EDX spectrum of NiO/CuO composite were captured using scanning electron microscope (TESCAN-Vega 3, Czech). The structural properties of resultant fibers and composite were identified by X-ray diffraction (XRD) technique (Bruker D8 Advance Diffractometer). The Electrochemical performance of the modified electrode were examined using a potentiostat instrument (Eco-Chemi B. V, Metrohm) and a three-electrode cell consisting of Ag/AgCl as a reference electrode, Pt wire as a counter electrode and NiO/CuO/GCE as a working electrode. The Cyclic voltammetry studies have been conducted in the potential range of -0.3 to 1 V and at the scanning rate of 20 $mVs^{-1}$.

**Results and discussion**

*X-ray diffraction*

The XRD patterns for both individual fibers and NiO/CuO composite are shown in Figure 1. In the NiO pattern, there are five peaks related to face-centered cubic structure of nickel oxide and no characteristic peaks of impurities can be detected. Diffraction peaks of CuO sample demonstrated in Figure1, match well with the monoclinic structure of CuO. The NiO/CuO pattern includes the peaks for both NiO and CuO planes, indicating the formation of NiO/CuO composite.

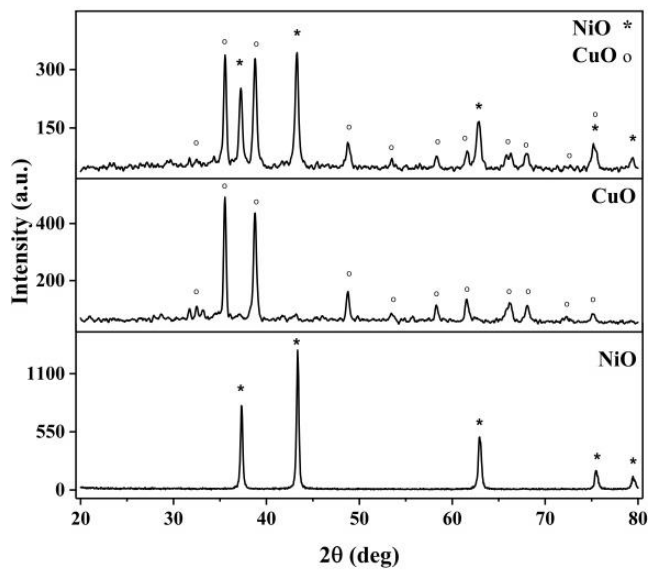

Figure. 1. XRD pattern for NiO, CuO and NiO/CuO nanocomposite.

*Microscopic image analyses*

The uniform and perfect fibrous morphology achieved by optimizing the spinning process parameters (i.e. voltage and flow rate). During calcination, the organic constituents belong to PVA, acetate and nitrate groups and, other volatiles (e.g. $H_2O$) have been decomposed. The average diameter of composite fibers before calcination was 270 nm and it decreased to 98 nm after calcination due to the break down of PVA and other volatile substances. In addition, the roughness of nanofibers' surface has been increased as a result of PVA elimination. The SEM images of electrospun composite fibers before and after calcination are shown in Figure 2.

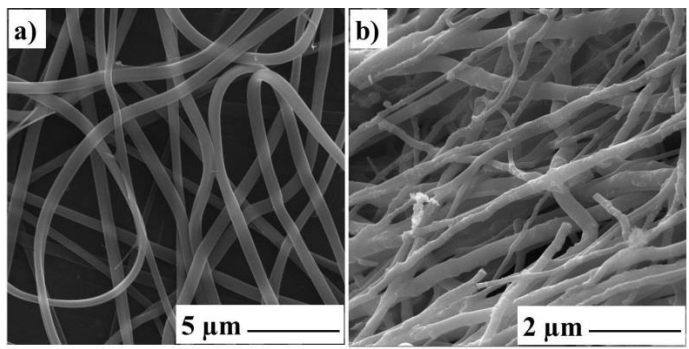

Figure. 2. SEM micrographs taken from PVA/ Nickel acetate/ Copper nitrate elecrospun fibers (a) and NiO/CuO nanofibers (b).

In order to observe the final morphology of the NiO/CuO/nafion ink after coating, a drop of the test solution was coated on an alumina substrate. SEM images were obtained from the top and cross section of the layers. Figure 3 (a) indicates that the uniform layer with well dispersed nanofibers has been produced by spin coating method. The cross sectional view of the deposited layer reveals that the layer thickness was about 6 μm and because of ultrasonic dispersion, the nanofibers have been fragmentized into shorter filaments, Figure 3 (b and c).

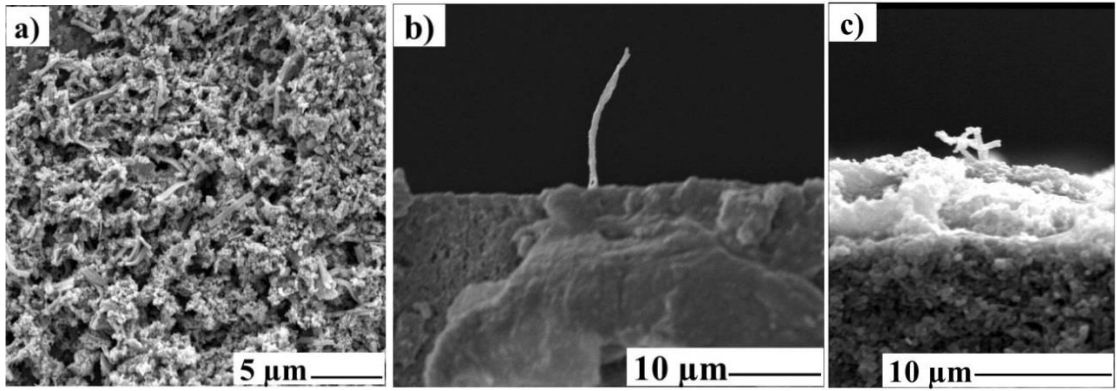

Figure. 3. SEM micrographs taken from the surface (a) and cross section of NiO/CuO/Nafion ink coated on the alumina substrate (b) and (c).

EDX spectrum of the NiO/CuO nanofibers is depicted in Figure 4. The peaks of Nickel, Copper and Oxygen atoms were detected. From the spectroscopy result, the molar ratio of NiO to CuO in the final produced composite was calculated to be 3.22. The neglectable amount of Al in the spectrum is related to the Aluminum foil that is used to collect the electrospun fibers.

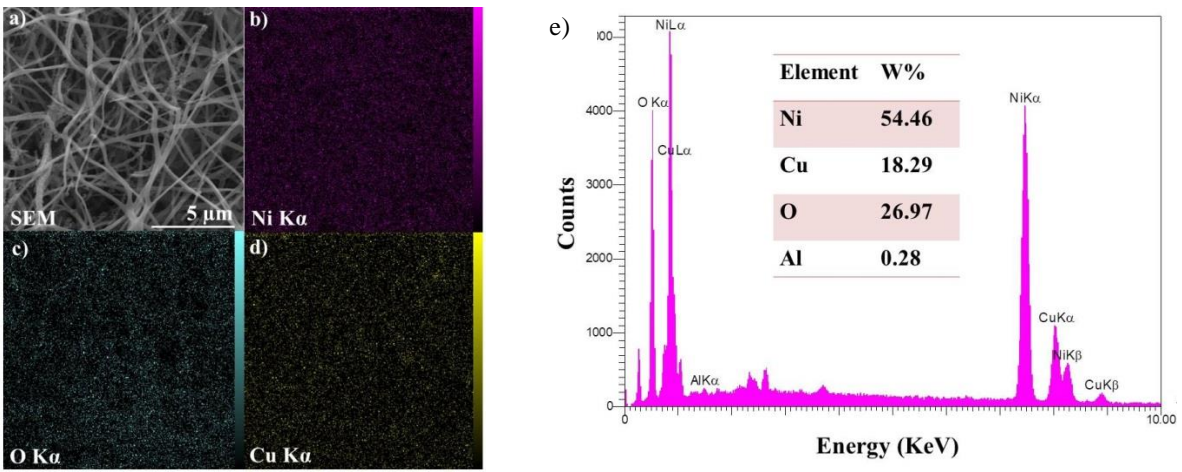

Figure. 4. (a) SEM image, (b-d) EDS spectrum , and (e) EDS elemental map for NiO/CuO nanocomposite.

*Electrochemical studies*

The comparative examinations on the electrocatalytic performance of bare GCE, NiO/GCE and NiO/CuO/GCE toward cholesterol detection were conducted by cyclic voltammetry (CV). The bare glassy carbon electrode that is depicted in Figure 5, implies no redox response to the presence of cholesterol. By adding of 0.8 mM cholesterol to the PBS solution, the CV curve of NiO/GCE shows a limited response which is attributed to the electrocatalytic activity of NiO and is consistent with previous reports [24, 31]. Whereas, the shape of CV for the NiO/CuO/GCE has changed drastically in the given potential range. The increase of oxidation current indicates synergic effect of NiO and CuO towards the oxidation of cholesterol.

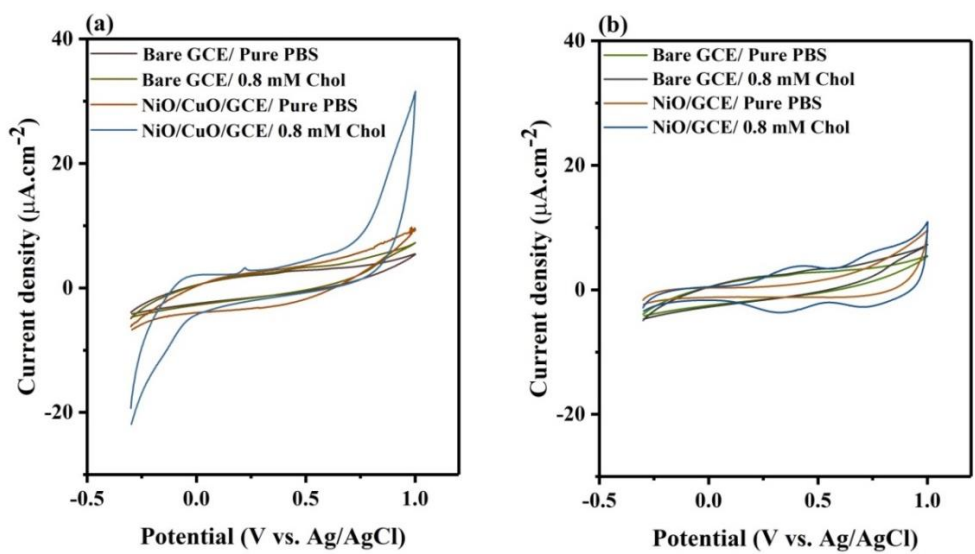

Figure. 5. Cyclic voltammmograms of NiO/CuO/GCE (a) and NiO/GCE (b) compared to bare electrode function in the absence and presence of 0.8 mM cholesterol.

The cyclic voltammograms of NiO/CuO/GCE and NiO/GCE in 0.1 M PBS containing different concentrations of cholesterol are illustrated in Figure 6. In both of these cases, an increase in response currents is accompanied with the increment of the cholesterol concentration. However, NiO/CuO electrode shows a more intense respond. Upon addition of cholesterol, a well-defined anodic peak with the onset potential at 0.9 V appeared. As the potential scanned in the negative direction (cathodic sweep), a reduction peak at -0.14 V was observed. This trend is depicted in Figure 6 (a). It can be concluded that the modified electrode with NiO/CuO nanofibers displays bi-directional electocatalytic ability toward the oxidation/reduction of cholesterol. The enlargement of anodic and cathodic peak currents of NiO/CuO/GCE is attributed to the presence of CuO and can be explained as follow. The water molecules are adsorbed on the surface of NiO nanostructure, transform it to the hydroxide (i. e. $Ni(OH)_2$). The $Ni(OH)_2$ is oxidized at the electrode surface and produces $Ni^{3+}$ species as NiOOH. Meanwhile, the NiOOH oxidize cholesterol to cholestenon and reduces to $Ni(OH)_2$ specie again. This mechanism is expressed in the equations 3 and 4 [24, 32]:

$$Ni(OH)_2 + OH^- \rightarrow NiOOH + H_2O + e^-. \tag{Eq. 3}$$

$$NiOOH + Cholesterol \rightarrow Cholestenone + Ni(OH)_2. \tag{Eq. 4}$$

The number of $Ni(OH)_2$ species that are generated spontaneously is limited. It can be raised by either providing an alkaline environment or doping a cation with a lower valency [32, 33]. Although the increase of pH value facilitate the oxidation of cholesterol and increase the current response of biosensor [2], the high pH of alkaline medium cannot mimic the biological condition and does not lead to a reasonable result. By introducing a cation such as CuO, the concentration of $Ni^{2+}/Ni^{3+}$ redox pair increases according to the reaction in the equation 5 [33]:

$$2Cu^+ + 4Ni^{2+} + 1/2O_2 \rightarrow 2Cu^+(Ni^{2+}) + 2Ni^{3+} + O^{2-}. \tag{Eq. 5}$$

The generation of charge careers ($Ni^{3+}$) results in the significant increase in electrical conductivity of nanocomposite. Also, Cholesterol can be directly oxidized by Cu(III) species. The oxidation of cholesterol probably occurs through the following reactions [31]:

$$CuO + OH^- \rightarrow CuOOH + e^-. \tag{Eq. 6}$$

$$CuOOH + Cholesterol \rightarrow Cholestenone + CuO + OH^-. \tag{Eq. 7}$$

Therefore, it leads to the conclusion that the higher oxidation current in the composite sample is attributed to the synergistic effect of redox couples of Ni(III)/Ni(II) and Cu(III)/Cu(II).

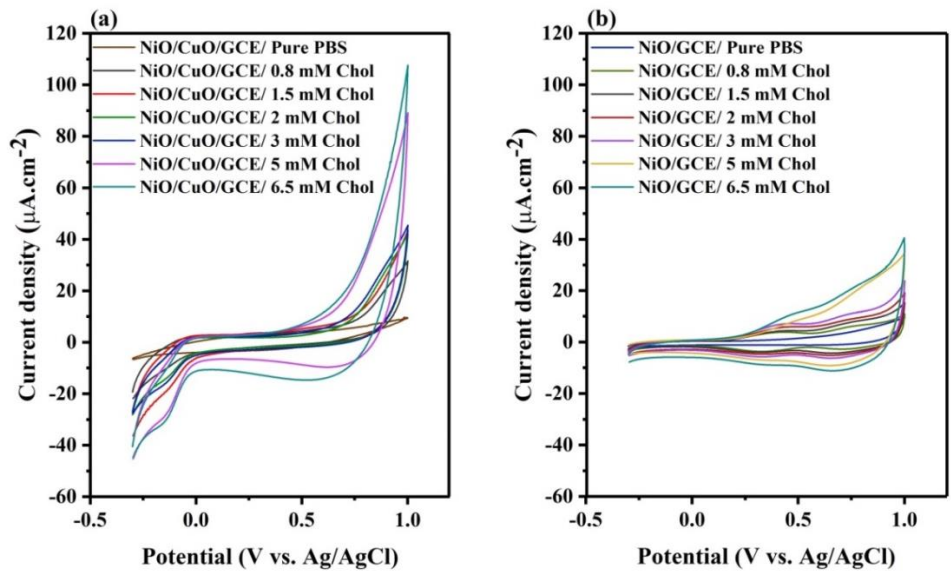

Figure. 6. Cyclic voltammograms of NiO/CuO/GCE (a) and NiO/GCE (b) at different cholesterol concentrations in 0.1 M PBS.

The sensitivity of the modified electrodes was calculated from the calibration plot for peak oxidation current as a function of cholesterol concentration, see Figure 7. The NiO/CuO/GCE has a good sensitivity of 10.27 $\mu A.mM^{-1}.cm^{-2}$. This sensitivity is higher than the non enzymatic cholesterol sensor, based on porous tubular Ag nanoparticles (approximately 0.7 $\mu A.mM^{-1}.cm^{-2}$) [2] and less than enzyme free cholesterol sensor based on NiO/ graphene (40.6 $mA.mM^{-1}.cm^{-2}$) [24]. The glassy carbon electrode modified with nanocomposite has a linear detection range from 0.8 to 6.5 mM (r = 0.992) with the detection limit of 5.9 µM. The high sensitivity of the nanocomposite is attributed to its structure. Nano fibrous structure with high surface to volume ratio enables more target analytes to be absorbed to the active sites at the same time that enhances the sensitivity of the bioelectrode. The NiO/GCE found to have a linear range with Pearson correlation of 0.994. The sensitivity of the electrode modified with NiO nanofibers calculated to be 2.5 $\mu A.mM^{-1}.cm^{-2}$.

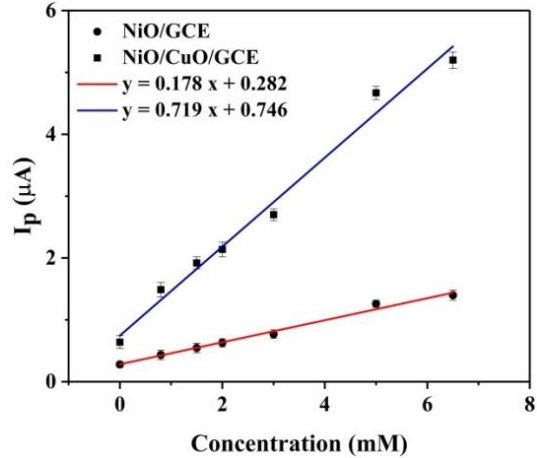

Figure. 7.Variation of anodic peak current with cholesterol concentration at the NiO/CuO/GCE obtained from the data in Figure 6(a).

To further study the electrochemical features of modified electrodes, chronoamperometry measurement for the modified electrode was recorded at the anodic peak potentials read from their CV plots (0.9 V for NiO/CuO/GCE and 0.75 V for NiO/GCE). A 50 μl 0.28 M cholesterol solution was injected successively to the 50 ml 0.1 M stirring PBS and the current was measured over time. As shown in Figure. 8, upon addition of cholesterol, the response current rises slightly. The NiO/CuO/GCE shows the response time of 3 seconds while the peaks for NiO/GCE reach to their climax in 2 seconds. On the other hand, the amperometric peaks for the electrode modified with the nanocomposite appeared earlier than the NiO/GC electrode.

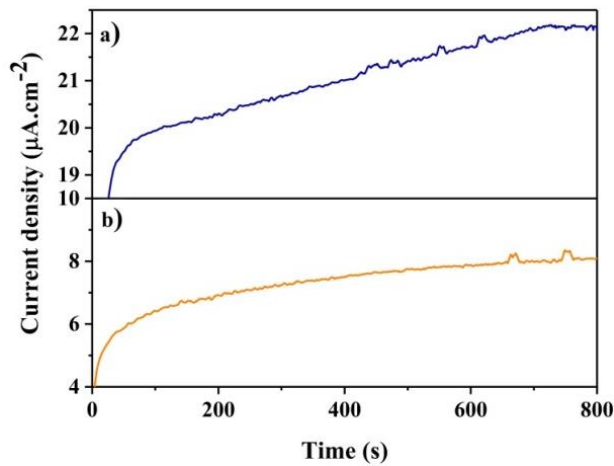

Figure. 8. Chronoamperometric response of (a) NiO/CuO/GCE and (b) NiO/GCE toward successive addition of 50 µl 0.28M cholesterol at an applied potential of 0.9 and 0.75 V, respectively.

The presence of electroactive species in blood may affect biosensor operation toward cholesterol detection. In order to examine the NiO/CuO/GC biosensor selectivity, some oxidizable compounds such as urea (2.3 mM), ascorbic acid (55µM), tyrosine (70µM), glucose (5mM) and glycine (250µM) were added at their normal physiological level to the 0.1 M, stirring PBS. Figure 9 represents the amperometric responses of interfering species and cholesterol at the NiO/CuO/GCE. It has been observed that there was no appreciable change in response current after adding any of abovementioned substances. However, upon addition of 4mM cholesterol, a well-defined current response was observed. The results revealed that co-existing compounds do not disturb the electrocatalysis current response of modified electrode.

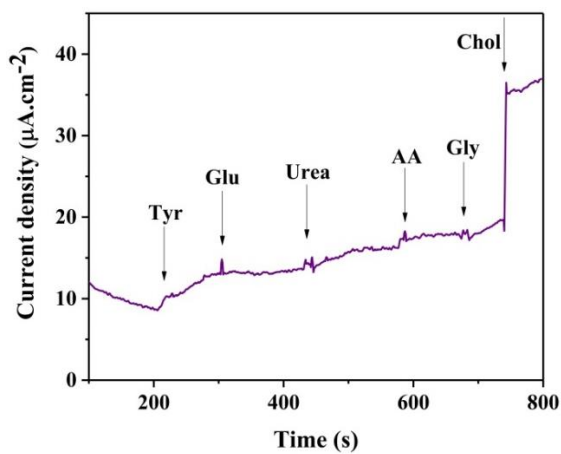

Figure. 9. Current response of NiO/CuO/GCE electrode to interfering species at an applied potential of 0.9 V.

To investigate the shelf life of the produced bioelectrodes, CV studies were performed at 5 mM cholesterol solution, in 2-day intervals, over 23 days and the magnitude of peak current was recorded each time. The electrode was washed with distilled water after every test and kept under the atmospheric condition. The comparison of the response current value at successive days with its initial value indicates that the bioelectrode that is modified with the nanocomposite, has a good stability retaining 81 % of activity after more than 3 weeks. The NiO/GCE showed a loss of 46 % in current response after 12 cycles. Figure 10 shows the response current density adapted from the cyclic voltamogramms of two modified electrodes in the span of 12 test cycles. The remarkable stability of NiO/CuO nanocomposite is attributed to the presence of CuO.

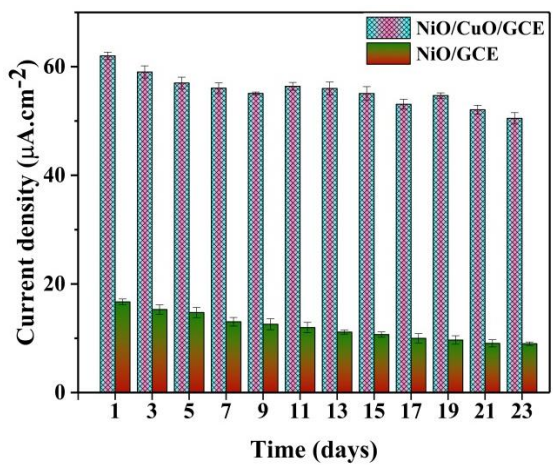

Figure. 10. Long term stability of cholesterol sensors.

**Conclusion**

In the current study, a non-enzymatic cholesterol biosensor based on Nickel oxide/Copper oxide nanocomposite is presented. The NiO/CuO nanocomposite was prepared by electrospinning method combined with appropriate calcination. The structural and morphological information of the nanocomposites were evaluated through different characterization techniques. The sensor exhibited high sensitivity, fast response time, long term stability and good selectivity which in addition to economic production route makes it a capable candidate for the industrial production. The net NiO fibrous thin film was examined as electrode modifier as well. Although NiO nanostructure has a good electrocatalytic behavior, it displayed far lower electroactivity, toward cholesterol detection compared to nanocomposite, because of its poor conductivity. The presence of Copper oxide has drastically enhanced the concentration of charge carriers in the nanocomposite sensor. In this study, the ratio of NiO to CuO was about 3 to 1. The authors suggest that the other proportions of Nickel to Copper oxide be considered in order to assess the optimized composition for higher sensitivity.